\documentclass[10pt]{iopart}
\usepackage{iopams,gensymb,amssymb,graphicx,xcolor}  

\begin{document}

\title{Magnetic phase diagram of a 2-dimensional triangular lattice antiferromagnet Na$_2$BaMn(PO$_4$)$_2$}

\author{Jaewook Kim$^1$, Kyoo Kim$^1$, Eunsang Choi$^2$, Young Joon Ko$^3$, Dong Woo Lee$^1$, Sang Ho Lim$^1$, Jong Hoon Jung$^3$, and Seungsu Lee$^1$}

\address{$^1$ Korea Atomic Energy Research Institute, Daejeon 34057, Republic of Korea}
\address{$^2$ National High Magnetic Field Laboratory, Florida State University, Tallahassee, FL 32310-3706, U.S.A.}
\address{$^3$ Department of Physics, Inha University, Incheon 22212, Republic of Korea}

\ead{jaewook@kaeri.re.kr}
\vspace{10pt}
\begin{indented}
\item[]August 2022
\end{indented}

\begin{abstract}
We report the magnetic phase transitions of a spin-5/2, 2-dimensional triangular lattice antiferromagnet (AFM) Na$_2$BaMn(PO$_4$)$_2$.
From specific heat measurements, we observe two magnetic transitions at temperatures 1.15~K and 1.30~K at zero magnetic field.
Detailed AC magnetic susceptibility measurements reveal multiple phases including the $\uparrow$$\uparrow$$\downarrow$ (up-up-down)-phase between 1.9~T and 2.9~T at 47~mK when magnetic field is applied along the $c$ axis, implying that Na$_2$BaMn(PO$_4$)$_2$ is a classical 2$d$ TL Heisenberg AFM with easy-axis anisotropy.
However, it deviates from an ideal model as evidenced by a hump region with hysteresis between the $\uparrow$$\uparrow$$\downarrow$ and $V$-phases and weak phase transitions.
Our work provides another experimental example to study frustrated magnetism in 2$d$ TL AFM which also serves as a reference to understand the possible quantum spin liquid behavior and anomalous phase diagrams observed in sibling systems Na$_2$Ba$M$(PO$_4$)$_2$  ($M$~=~Co, Ni).
\end{abstract}

%
\vspace{2pc}
\ioptwocol

\section{Introduction}

The 2-dimensional triangular lattice (2$d$ TL) antiferromagnetic (AFM) system has served as a cornerstone in the field of geometrical frustrated magnetism \cite{Collins_RN1353,Ramirez_RN1273,Starykh_RN1367}.
Pioneering works on Ising AFM system showed that it maintains a disordered state with sizable residual entropy down to zero temperature ($T$) \cite{Wannier_RN1316,Husimi_RN1337}. 
The idea of ``resonating valence bond (RVB)'' has been proposed on Heisenberg system with spin $S$~=~1/2 by Anderson \cite{Anderson_RN1368}, where magnetic long-range-order (LRO) is suppressed by quantum fluctuations.
Although later works have shown that its ground state is similar to its classical-spin counterpart, an ordered state with a coplanar 120~$^{\circ}$ phase \cite{Bernu_RN1421, Capriotti_RN2135,White_RN1506}, the RVB model has evolved to the concept of ``quantum spin liquid'' (QSL) \cite{Balents_RN1917,Mila_RN1431,Savary_RN1472,Zhou_RN1537}, which led to researches on geometrically frustrated systems such as pyrochlore and kagome lattice systems \cite{Bramwell_RN1398,Gardner_RN1449,Norman_RN1482}.

\begin{figure}
\centering
\includegraphics[width=\columnwidth]{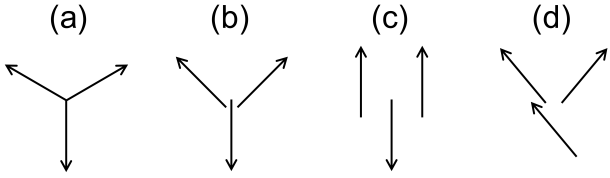}
\caption{
Schematic representations of the (a) 120~$^{\circ}$, (b) $Y$, (c) up-up-down ($\uparrow$$\uparrow$$\downarrow$), and (d) $V$-phases.
}
\label{Fig1}
\end{figure}

The evolution of the ordered phase of 2$d$ TL Heisenberg AFM under temperature ($T$) and magnetic field ($H$) has been extensively studied theoretically.
Under application of magnetic field, the coplanar 120~$^{\circ}$ phase, which is the $H$~=~0 ground state of both quantum ($S$~=~1/2) and classical ($S$~$\rightarrow$~$\infty$) spin systems, evolves to the $Y$, $\uparrow$$\uparrow$$\downarrow$ (up-up-down), and $V$-phases (see Fig.~\ref{Fig1}) upon increasing the magnetic field, stabilized by quantum and thermal fluctuation, respectively \cite{Chubukov_RN2252,Kawamura_RN2239,Seabra_RN1350,Gvozdikova_RN2238}.
Considering the classical Heisenberg spin case, at $T$~=~0, the $\uparrow$$\uparrow$$\downarrow$-phase is stable only at the 1/3 of saturation magnetic field but expands to a finite range of magnetic field at non-zero $T$ and exhibits the highest transition temperature among other phases.
At zero magnetic field, earlier studies suggested that the system would undergo a single phase transition from paramagnetic (PM) to the coplanar 120~$^{\circ}$ phase \cite{Kawamura_RN2239}.
However, recent numerical works have shown that there are actually two transitions when magnetic field is sufficiently close to zero; from high $T$ PM phase to $\uparrow$$\uparrow$$\downarrow$ and then to $Y$-phase upon cooling \cite{Gvozdikova_RN2238,Seabra_RN1350}, each transitions corresponding to different symmetry breakings.

Surprisingly, there are only a few systems that realize the classical 2$d$ TL Heisenberg AFM model discussed above, despite the fact that its simple structural motif is realized in many materials \cite{Kitazawa_RN2215,Svistov_RN2201,Ishii_RN1386,Zhou_RN1657,Quirion_RN2269}.
Moreover, systems with magnetically easy-axis anisotropy are scarcer compared to easy-plane ones: 
as far as we know, only GdPd$_2$Al$_3$ \cite{Kitazawa_RN2215} and Rb$_4$Mn(MoO$_4$)$_3$ \cite{Ishii_RN1386} fall into this category.
In practice, various perturbations, such as anisotropic \cite{Alicea_RN2256}, next-nearest-neighbor (NNN) \cite{Takagi_RN2257,Swanson_RN2258}, and nearest-neighbor (NN) interlayer \cite{Yamamoto_RN2261} exchange interactions intervene and modify the ideal/simple ground state.

Recently, Na$_2$Ba$M$(PO$_4$)$_2$ system (NB$M$PO, $M$~=~Co, Ni) was reported to show frustrated magnetism within the 2$d$ TL structure.
NBCoPO was proposed as a candidate for QSL since it consists of Co$^{2+}$ ions with effective spin-1/2.
First experimental work demonstrated that NBCoPO is magnetically disordered down to 0.05~K with a broad continuum in inelastic neutron scattering, probably due to spinon excitations \cite{Zhong_RN1618}.
However, another study showed that an AFM phase transition occurs at the N\'{e}el temperature ($T_N$) of 0.148~K in $H$~=~0 in the same compound \cite{Li_RN2099}.
Although this observation excludes the QSL as its ground state, thermal conductivity ($\kappa$) above $T_N$ follows a $T$-linear behavior and exhibits a non-zero value of $\kappa$/$T$ at the $T$~=~0 intercept, implying that there are itinerant thermal carriers, also possibly due to spinons \cite{Li_RN2099}.
Another system, NBNiPO, which has a slightly distorted structure but maintains the 2$d$ TL structure, displays a phase transition at 0.43~K in $H$~=~0 \cite{Ding_RN2105,Li_RN2117}.
Both systems mentioned above show successive phase transitions under applied magnetic field, including the $\uparrow$$\uparrow$$\downarrow$-phase \cite{Li_RN2099,Li_RN2117}.

Here, we report the physical properties and magnetic phase diagram of Na$_2$BaMn(PO$_4$)$_2$ with spin-5/2 of Mn$^{2+}$ ions, realizing the classical 2$d$ TL Heisenberg AFM in the NB$M$PO series.
Low temperature specific heat ($C_p$) measurement shows two phase transitions at 1.15 and 1.30~K under $H$~=~0.
Entropy analysis suggests that almost half of the measured entropy is released above the ordering temperature, suggesting strong magnetic frustration persists in the PM phase.
Combining the $C_p$ and AC magnetic susceptibility data, we map the complete $H$~--~$T$ phase diagram of this material.
When magnetic field is applied along the $c$ axis, we find three magnetic field-induced phases ($Y$, $\uparrow$$\uparrow$$\downarrow$, and $V$) that are in good agreement with those expected for a classical 2$d$ TL Heisenberg AFM model with easy-axis anisotropy.
Surprisingly, we find an unusually wide hysteresis hump of $\Delta$$H$~$\sim$~0.7~T between the $\uparrow$$\uparrow$$\downarrow$ and $V$-phase.
While both phase boundaries of the $\uparrow$$\uparrow$$\downarrow$-phase in the model 2$d$ TL Heisenberg AFM is known to be a continuous transition of the Berezinskii-Kosterlitz-Thouless (BKT) type, observation of hysteresis in NBMnPO suggests that the system involves additional exchange interactions.
Our work adds a rare example in classical 2$d$ TL Heisenberg AFM systems with easy-axis anisotropy which can be fully mapped in a typical laboratory magnetic field scale and serves as a reference to understand quantum phenomena and complex phase diagram that are observed in other NB$M$PO systems.

\section{Methods}
\begin{figure}
\centering
\includegraphics[width=\columnwidth]{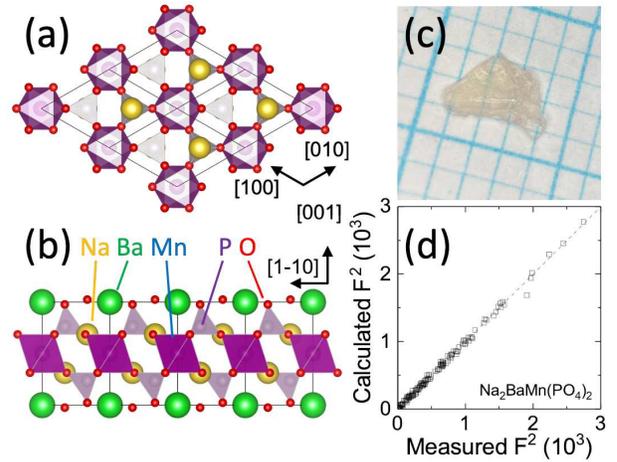}
\caption{
Crystal structure of Na$_{2}$BaMn(PO$_4$)$_2$ viewed along (a) [001] and (b) [110] direction.
Na, Ba, Mn, P, and O ions are shown in yellow, green, blue, violet, and red spheres, respectively.
Mn and P ions are enclosed in oxygen octahedra and tetrahedra, respectively.
(c) Photograph of a single crystal on 1~$\times$~1 mm$^2$ scale grid.
(d) Calculated structure factor squared ($F^2$) versus measured $F^2$ according to the single crystal XRD fitting.
Dashed line is a guide to the eye.
}
\label{Fig2}
\end{figure}

Single crystals of NB$M$PO ($M$~=~Mg, Mn, Co, and Ni) were grown by high temperature flux method similar to a previous report \cite{Zhong_RN1618}.
Typical dimension of the samples is $\sim$~1.0~--~3.0~mm in edge length and $\sim$~0.3~mm in thickness.
NBMnPO single crystals are transparent in color, as shown in Fig.~\ref{Fig2}(c), and are planar in shape with the planes perpendicular to the $c$ axis.
Single crystal X-ray diffraction (XRD) data were collected by using graphite-monochromated Mo K$\alpha$ radiation (PHOTON II X-ray diffractometer, Bruker) and were refined with SHELXL-13 \cite{Sheldrick_RN2276}.
DC magnetic susceptibility ($\chi$) and magnetization ($M$) were measured using a superconducting quantum interferometer (Magnetic Properties Measurement System, Quantum design).
Specific heat ($C_p$) was measured by using a relaxation method in a cryostat equipped with a superconducting magnet (Dynacool-14, Quantum Design).
A dilution refrigerator (DR) was used to access low temperature down to 0.08~K.
AC magnetic susceptibility measurements were performed by using a homemade susceptometer with a voltage controlled current source (CS580, Stanford Research) and lock-in amplifier (SR830, Stanford Research).
The phase of the lock-in amplifier is set to measure the first harmonic signal.
We used the SCM1 magnet equipped with a DR in National High Magnetic Field Laboratory (Tallahassee, FL, U.S.A.).

\section{Results}
NBMnPO was first reported as a naturally occurring mineral with space group $P\bar{3}$ \cite{Nishio-Hamane_RN2104}.
A subsequent study, using a chemically pure powder form, showed that its actual space group is $P\bar{3}m1$ \cite{Nenert_RN2103}.
Our single crystal XRD refinement result (Fig.~\ref{Fig2}(d)) is consistent with space group $P\bar{3}m1$ with lattice parameters $a$~=~5.3761(9)~\AA~and $c$~~=~7.0999(12)~\AA (see Supplemental Material \cite{Supple} for best fitting results and structural parameters) and previous reports including other isostructural NB$M$PO systems ($M$~=~Co and Mg) \cite{Boukhris_RN2098,Zhong_RN1618}.
As shown in Fig.~\ref{Fig2}(a,b), Mn$^{2+}$ ions are enclosed in perfect $M$O$_6$ octahedra connected by distorted PO$_4$ tetrahedra, forming equilateral triangular layers.
Mn$^{2+}$ ion is a half-filled 3$d^{5}$ electron system with high spin-5/2 and zero angular momentum, representing a Heisenberg spin.
These layers are stacked along the $c$ axis, separated by Ba ions.
The dominant interaction between these spins are presumably the NN intralayer superexchange interaction through Mn--O--P--O--Mn.

\begin{figure}
\centering
\includegraphics[width=0.7\columnwidth]{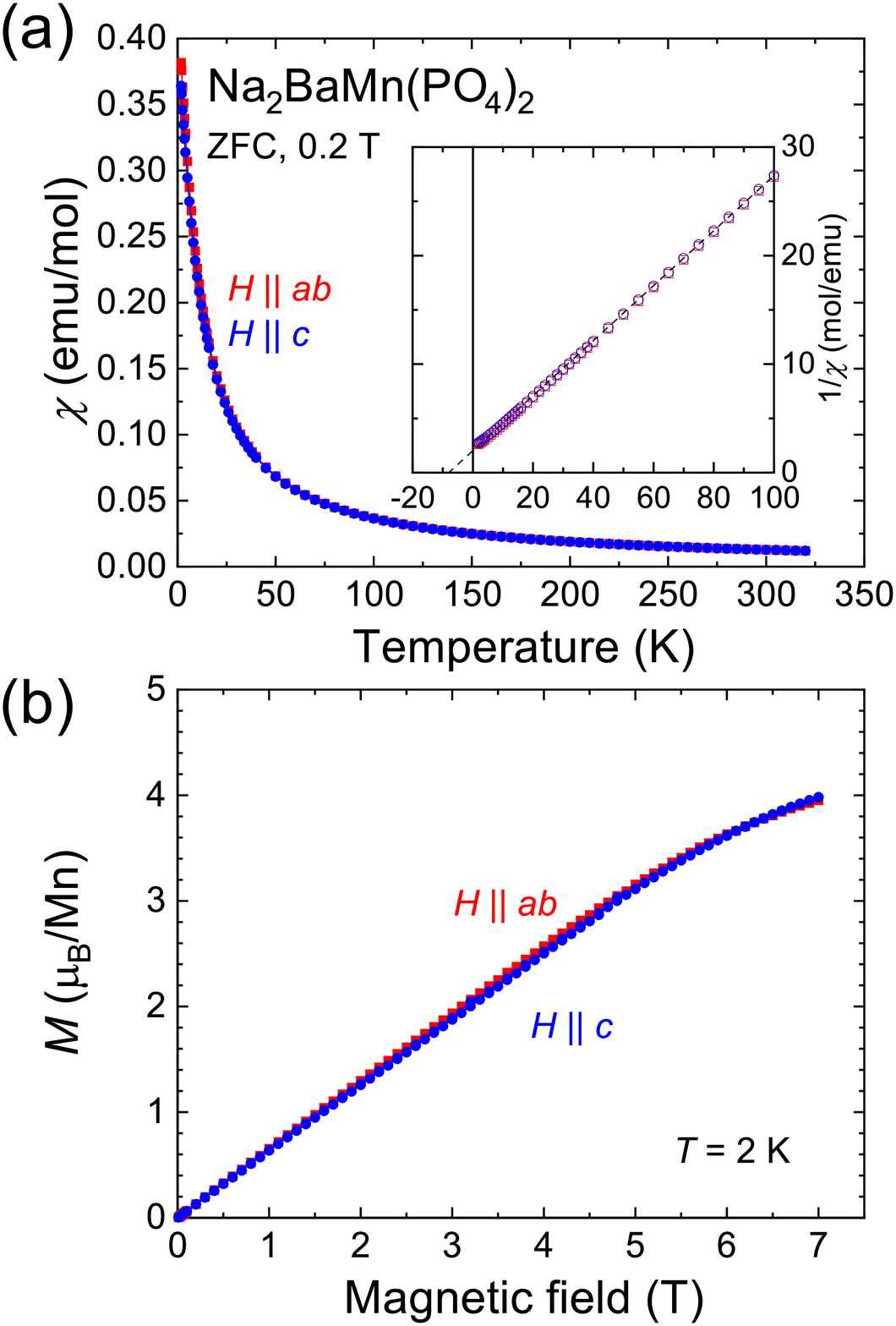}
\caption{
Magnetic properties of Na$_2$BaMn(PO$_4$)$_2$ single crystal.
(a) Temperature-dependences of magnetic susceptibility measured along two crystallographic directions.
(b) Magnetic field-dependences of magnetization at 2~K.
Red circle and blue square symbols represent the data measured along $ab$ plane and $c$ axis, respectively.
Inset in (a) shows the inverse of magnetic susceptibility as a function of temperature.
Dashed line is a guide to the eye.
}
\label{Fig3} 
\end{figure}

Next, we discuss the magnetic properties of NBMnPO single crystal.
Temperature-dependence of DC magnetic susceptibility data were taken while warming under $\mu_0$$H$~=~0.2~T after zero field cooling.
$\chi$($T$) curves show monotonic increase upon cooling from 320 to 1.8~K without any signature of LRO (Fig.~\ref{Fig3}(a)).
The system is more-or-less magnetically isotropic down to 1.8~K with anisotropy ratio ($\equiv$~$\chi_{\parallel}$$_{c}$/$\chi_{\parallel}$$_{ab}$) of 0.95 at 2~K.
We applied the Curie-Weiss (CW) model at 20~K~$\leq$~$T$~$\leq$~100~K range and found that the CW temperature ($\Theta_{CW}$) and effective magnetic moment ($\mu_{eff.}$) values of -7.20~K (-7.61~K) and 5.61~$\mu_B$/Mn (5.60~$\mu_B$/Mn) for $H$~$\parallel$~$c$ ($H$~$\parallel$~$ab$), respectively.
The $\mu_{eff.}$ values are slightly smaller than that expected for spin-only Mn$^{2+}$ (high-spin, 5.92~$\mu_B$).
Magnetization measurements at 2~K also show isotropic behavior for both directions (Fig.~\ref{Fig3}(b)).
The slope changes above $\sim$~6~T, but does not saturate up to the maximum magnetic field measured.

The frustration ratio ($\equiv$~$|\Theta_{CW}|$/$T_N$) of NBMnPO is 5.85, similar to other classical 2$d$ TL Heisenberg AFM systems, $cf.$ Rb$_4$Mn(MoO$_4$)$_3$ (7.14) \cite{Ishii_RN1386}, RbFe(MoO$_4$)$_2$ (5.79) \cite{Svistov_RN2201}, and NBNiPO (4.3) \cite{Li_RN2117}, signaling a moderate magnetic frustration.
Assuming the mean-field theory, the NN intralayer exchange interaction $J$ of the Heisenberg Hamiltonian $\mathcal{H}$~=~$\sum_{i,j} J \vec{S_i} \cdot \vec{S_j} $ can be estimated from the equation $\Theta_{CW}$~=~$-z J S(S+1)/3 k_B$, where $z$ is the number of nearest-neighbors.
For NBMnPO, we find $J$~$\sim$~0.43~K which is comparable to that of NBNiPO (0.48~K) \cite{Li_RN2117}.

\begin{figure}
\centering
\includegraphics[width=1.04\columnwidth]{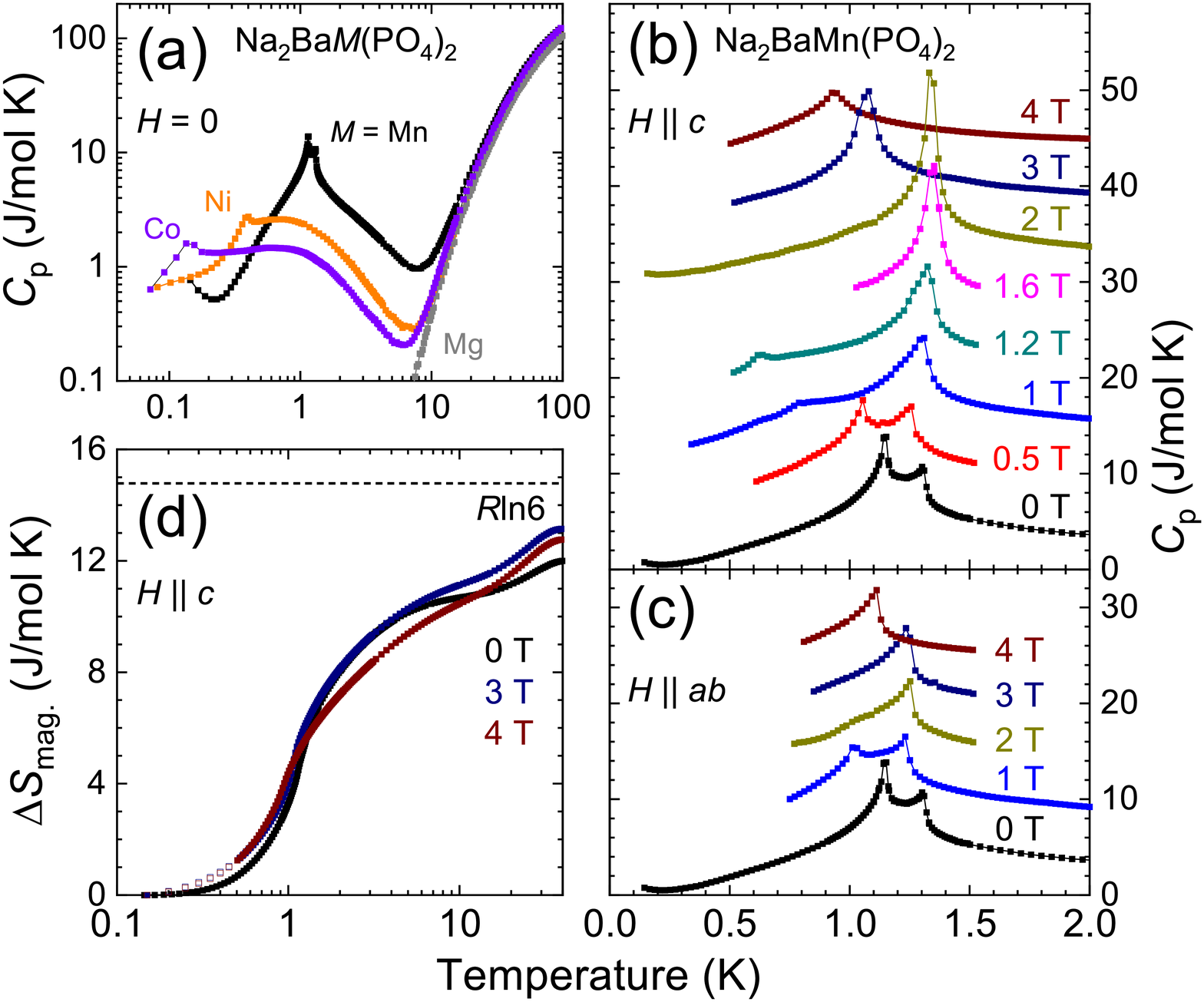}
\caption{
(a) Specific heat of different Na$_2$Ba$M$(PO$_4$)$_2$ ($M$~=~Mg, Mn, Co, and Ni) single crystals as a function of temperature in log-log scale measured at zero magnetic field.
Specific heat of Na$_2$BaMn(PO$_4$)$_2$ as a function of temperature below 2~K under different magnetic fields applied along the (b) $c$ axis and (c) $ab$ plane.
All curves in (b) ((c)), except for $H$~=~0, are sequentially shifted by 6~J/mol~K (5.5~J/mol~K) for clarity.
(d) Change of magnetic entropy ($\Delta$$S_{mag.}$) as a function of temperature in semi-log scale.
Dashed line in (d) indicates the $R$ln6 value as explained in text.
Open squares for $\mu_0$$H$~=~3, 4~T below 0.5~K, are extrapolated data that follow a power law fit below $T_N$, for comparison with the 0~T curve.
}
\label{Fig4}
\end{figure}

To further study the phase transition at lower $T$, we measured the specific heat down to 0.08~K.
Fig.~\ref{Fig4}(a) shows the $C_p$($T$) curves measured for NB$M$PO single crystals ($M$~=~Mg, Mn, Co, and Ni).
NBMgPO is measured as a non-magnetic reference for lattice contribution of specific heat.
All three systems with magnetic ions show similar behaviors: upturns (i.e., local minimum) in $C_p$($T$) upon cooling below 6--8~K and peak features at low $T$ indicating magnetic LRO, consistent with previous reports \cite{Ding_RN2105,Li_RN2099,Li_RN2117}.
Among them, NBMnPO shows the highest upturn temperature at 7.7~K.

A closer look at the $C_p$($T$) data below 2~K reveals two phase transitions in NBMnPO (Fig.~\ref{Fig4}(b)) with a strong magnetic field-dependence.
At $H$~=~0, two sharp peaks appear at 1.30~K ($T_{N1}$) and 1.15~K ($T_{N2}$) indicative of magnetic phase transitions .
Under $H$~$\parallel$~$c$, the peak at $T_{N2}$ reduces concomitantly in size and shifts to lower $T$ and disappears at $\mu_{0}$$H$~$>$~2~T, while that of $T_{N1}$ initially shifts to low $T$ at 0.5~T then gradually moves to higher $T$, becomes sharpest at 2~T, and finally shifts to lower $T$ and reduces in size.
Another small upturn is observed below 0.23~K in $H$~=~0 curve, which originates from nuclear Schottky anomaly due to large nuclear spin of Mn ($I$~=~5/2).
When magnetic field is applied along the $ab$ plane (Fig.~\ref{Fig4}(c)), $T_{N1}$ decreases slightly at 1~T, stays nearly constant in $T$ up to 3~T and gradually shifts to lower $T$ above 4~T, whereas that of $T_{N2}$ shifts to 1.0~K at 1~T and shows a weakly smeared behavior at the same temperature at 2~T.

The magnetic specific heat $C_{mag.}$ of NBMnPO is obtained by subtracting the $C_p$ data of NBMgPO as phonon contribution and nuclear Schottky contribution ($\propto$~$T^{-2}$) at very low temperature from that of NBMnPO.
Then, $C_{mag.}$/$T$ was integrated as a function of $T$ from 0.15~K to obtain the magnetic entropy ($\Delta$$S_{mag.}$).
Fig.~\ref{Fig4}(d) displays the $\Delta$$S_{mag.}$($T$) in several magnetic fields under $H$~$\parallel$~$c$.
For comparison, data for 3 and 4~T between 0.5~K and 0.15~K were extrapolated (shown as open symbols) assuming a power-law dependence of $C_p$($T$) curve below $T_N$.
At $H$~=~0, $\Delta$$S_{mag.}$ saturates to 12.0~J/mol~K at 30~K, which is 80.5~\% of $R$ln(2$S$~+~1)~=~14.90~J/mol~K value that is expected for Mn$^{2+}$ spins with $S$~=~5/2.
The $\Delta$$S_{mag.}$($T$) curves increase abruptly at two points: at $T_{N1}$ and at $\sim$~20~K.
A large fraction of observed entropy (51.2~\% of $\Delta$$S_{mag.}$($T$~=~30~K)) is released in $T_{N1}$~$<$~$T$~$<$~30~K range suggesting that spin fluctuation persists up to temperature that is roughly 20 times larger than $T_{N1}$.

\begin{figure*}
\centering
\includegraphics[width=2\columnwidth]{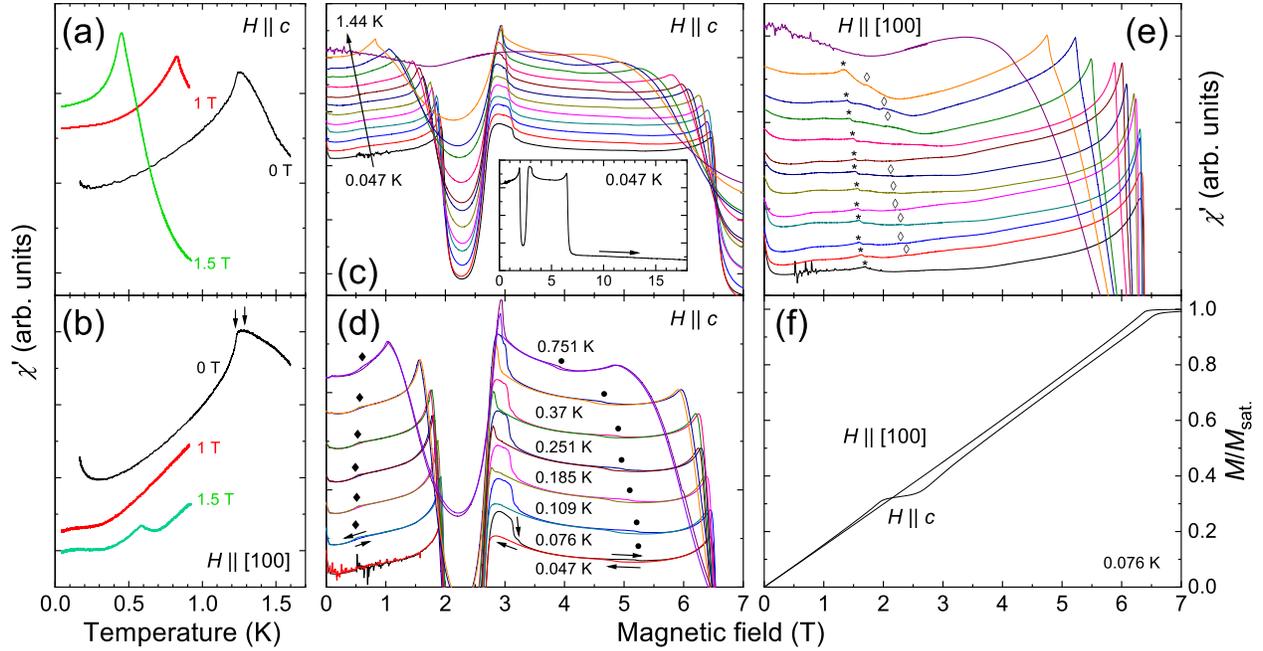}
\caption{
AC susceptibility and integrated magnetization of Na$_2$BaMn(PO$_4$)$_2$.
Temperature dependence of $\chi'$ measured with fixed magnetic field along the (a) $c$ axis and (b) $ab$ plane.
(c) Magnetic-field dependence of $\chi'$ measured with increasing magnetic field applied along the $c$ axis at constant temperatures.
Inset in (b) shows the AC susceptibility curve at 0.047~K measured up to 18~T.
(d) Enlarged graph of (b), showing hysteresis regions at different temperatures.
(e) Magnetic-field dependence curves measured with increasing magnetic field applied along the [100] direction at constant temperatures.
(f) Magnetization along two directions at 0.076~K obtained by integrating the AC susceptibility curves as described in text.
$\blacklozenge$, $\bullet$, *, and $\lozenge$ symbols indicate relatively weak transition features described in text.
Curves shown in (a--e) are shifted along the $y$ axis for clarity.
Temperatures for each curves in (c) and (e) are, from bottom to top, 0.047, 0.076, 0.109, 0.137, 0.185, 0.251, 0.342, 0.370, 0.480, 0.651, 0.751, 0.952, and 1.44~K, respectively.
Frequencies of AC magnetic field for the $c$ axis and $ab$ plane directions are 471~Hz and 717~Hz, respectively.
}
\label{Fig5}
\end{figure*}

To further track the phase transitions, we measured the low temperature AC magnetic susceptibility ($\chi'$) in detail.
Fig.~\ref{Fig5}(a) and (b) show the $\chi'$ vs. $T$ curves with both DC and AC magnetic field applied along the $c$ axis and [100] direction, respectively.
Both zero field curves show increase in $\chi'$ upon cooling, reaching a saturated behavior at 1.28~K and exhibits a sharp drop at 1.23~K (marked with black arrows in Fig.~\ref{Fig5}(b)), indicating the two phase transitions observed in $C_p$($T$).
Below 0.28~K, $\chi'$ increases slightly which is more pronounced in $H$~$\parallel$~$ab$ plane data.
$\chi'$($T$) data with applied magnetic fields were measured below $\sim$~0.9~K to preserve the $^3$He/$^4$He mixture at low temperature during the experiment.
In the $H$~$\parallel$~$c$ arrangement (Fig.~\ref{Fig5}(a)), a peak appears at 0.82~K at 1~T which shifts to lower $T$ at 1.5~T and in the $H$~$\parallel$~$ab$ arrangement (Fig.~\ref{Fig5}(b)), a small peak appears at 0.60~K at 1.5~T.

Fig.~\ref{Fig5}(c) shows the $\chi'$ versus $H$ curves with increasing the DC magnetic field along the $c$ axis.
At the lowest temperature, 0.047~K, we observe a sharp peak at $\mu_0$$H_{c1}$~=~1.9~T followed by a large dip, with its minimum at 2.3~T, which terminates as a broad maximum at $\mu_0$$H_{c2}$~=~2.9~T.
With further increase in magnetic field, $\chi'$ shows another peak at $\mu_0$$H_{c3}$~=~6.45~T and quickly decreases  and shows a linear behavior above $\mu_0$$H_{sat.}$~$\sim$~7.4~T (inset in Fig.~\ref{Fig5}(c)).
By increasing the temperature, the dip feature becomes broader and $H_{c1}$ shifts to lower magnetic field with broadening in the peak shape, while the one at $H_{c2}$ becomes sharper without significant change in peak position.

In Fig.~\ref{Fig5}(d), we show the enlarged view of $\chi'$($H$) curves shown in Fig.~\ref{Fig5}(c) with both up- and down-sweeps of magnetic field at selected temperatures.
There are three noticeable characteristics.
First, the broad maximum in $H_{c2}$ shows a strong hysteresis in magnetic field: in the up-sweep it manifests as a broad hump between 2.7 and  3.4~T ($T$~=~0.047~K), but it disappears in down-sweep and $\chi'$($H$) curve changes smoothly then enters the dip with a shoulder-like shape.
At higher temperatures, the hump observed in the up-sweep becomes narrow in $H$, and a peak develops above 0.251~K.
Second, a small shoulder (marked with $\blacklozenge$ in Fig.~\ref{Fig5}(d)) is observed below the $\uparrow$$\uparrow$$\downarrow$-phase which occur at 0.6~T (0.5~T) at 0.076~K in the up-sweep (down-sweep).
The position of this anomaly slightly shifts to higher magnetic field upon increase in temperature.
Third, another small shoulder ($\bullet$ in Fig.~\ref{Fig5}(d)) is observed above the $\uparrow$$\uparrow$$\downarrow$-phase, e.g., at 5.2~T (4.8~T) in the up-sweep (down-sweep) data at 0.076~K.
With increase in temperature, this feature quickly shifts to lower magnetic-field.

On the other hand, when magnetic field is applied along the $ab$ plane, $\chi'$($H$) curves show a simpler behavior compared to that of the $H$~$\parallel$~$c$ case (Fig.~\ref{Fig5}(e)).
For example, at $T$~=~0.076~K, three anomalies are observed; small peaks at $\mu_0$$H_{c4}$~=~1.7~T (* symbol), $\mu_0$$H_{c5}$~=~2.4~T ($\lozenge$ symbol), and a sharp peak at $\mu_0$$H_{c6}$~=~6.3~T.
Measurements under different $T$ show that the former one is almost independent in $T$ while the latter one shifts to lower $H$ as $T$ approaches $T_N$.

Since the shape of the $\chi'$($H$) curves are almost frequency-independent between 147~Hz to 991~Hz (See Supplementary Information \cite{Supple}), it can be referred to as the differential magnetic susceptibility (d$M$/d$H$) in the DC limit.
Then, by integrating the $\chi'$($H$) curves with respect to $H$ and assuming that the magnetization saturates above $H_{sat.}$, we obtain a normalized $M$/$M_{sat.}$($H$) curve at 0.076~K as shown in Fig.~\ref{Fig5}(f).
Changes of slope in $M$/$M_{sat.}$($H$) is clearly visible when $H$ is parallel to the $c$ axis, marking the 1/3 magnetization plateau with $M$/$M_{sat.}$~$\sim$~0.34 at 2.3~T, which behavior is expected for a $\uparrow$$\uparrow$$\downarrow$-phase.
This behavior also confirms that the system belongs to the easy-axis magnetic anisotropy.
On the contrary, when $H$ is applied along the $ab$ plane direction, magnetization increase linearly up to saturation magnetic field.

We note that, although the broad hump feature at $H_{c2}$ is clearly visible in the $\chi'$($H$) curves (Fig.~\ref{Fig5}(d)), its contribution to magnetization, i.e., the area of the hump between up- and down-sweep $\chi'$($H$) curves, is relatively small.
For example, at 0.076~K, the hump feature is $\sim$~3.1~\% of the $M_{sat.}$ and not apparent in the integrated $M$($H$) curve.

\begin{figure}
\centering
\includegraphics[width=0.75\columnwidth]{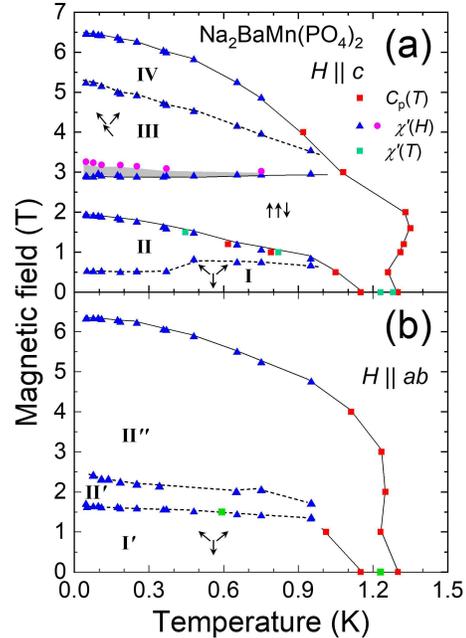}
\caption{
$H$~--~$T$ phase diagrams of Na$_2$BaMn(PO$_4$)$_2$ with magnetic field applied along the (a) $c$ axis and (b) $ab$ plane.
Red squares indicate the phase boundaries from peak in $C_p$($T$) data.
Blue triangles and magenta circles indicate anomalies observed in $\chi'$($H$).
Green squares indicate anomalies observed in $\chi'$($T$) data.
Grey area in (a) describes the hump region with hysteresis observed in AC susceptibility measurement as described in text.
Phase boundaries marked by solid lines are determined from peak positions in $C_p$($T$) and $\chi'$($H$) data.
Dashed lines are determined from weak anomalies in $\chi'$($H$).
}
\label{Fig6}
\end{figure}

Combining $C_p$($T$), $\chi'$($T$), and $\chi'$($H$) data, we obtain full $H$~--~$T$ phase diagrams of NBMnPO along two applied magnetic field directions as shown in Fig.~\ref{Fig6}.
When magnetic field is applied along the $c$ axis (Fig.~\ref{Fig6}(a)), we find a series of ordered phases denoted as I, II, $\uparrow$$\uparrow$$\downarrow$, III, and IV.
The peaks at $T_{N2}$ in $C_p$($T$) data smoothly connects to the peaks in $\chi'$($T$) at 1 and 1.5~T and $\chi'$($H$) at $H_{c1}$, constituting the lower field phase boundary of the $\uparrow$$\uparrow$$\downarrow$-phase while the peaks at $T_{N1}$ connects to the sharp peak features at $H_{c3}$, delineating the LRO to PM phase boundary (solid lines).
The expansion of $\uparrow$$\uparrow$$\downarrow$-phase is due to the thermal fluctuations \cite{Kawamura_RN2239,Seabra_RN1350,Gvozdikova_RN2238}, which is also observed 
in classical spin systems with other structure \cite{Yahne_RN2347}.
The higher field phase boundary of the $\uparrow$$\uparrow$$\downarrow$-phase is marked with $H_{c2}$ and is nearly linear in $H$~--$T$ phase diagram (solid line).
The region of the hump in the up-sweep of $\chi'$($H$), colored in grey, is located adjacent to this phase boundary on the high field side and terminates at the point where the hysteresis disappears (magenta circles).
The hysteresis feature persists up to 0.75~K.

When magnetic field is applied along the $ab$ plane (Fig.~\ref{Fig6}(b)), the phase diagram consists of three phases: I$'$, II$'$, and II$''$.
The $\uparrow$$\uparrow$$\downarrow$-phase is absent in this configuration, demonstrating the easy-axis anisotropy of this system.
The peaks at $T_{N1}$ in $C_p$($T$) curve shift to lower $T$ with increasing field, merging with the peaks at $H_{c6}$ observed in $\chi'$($H$).
The peak at $T_{N2}$ decreases quickly in magnetic field and seem to merge with the peak in $\chi'$($T$) at 1.5~T and weak $\chi'$ anomalies at $H_{c4}$.
Phase II$'$ lies in between the weak transition features observed at $H_{c4}$ and $H_{c5}$.

\noindent
\section{Discussion}

Having assigned the $\uparrow$$\uparrow$$\downarrow$-phase from the 1/3 magnetization plateau, we can assume that the low- (phase I, II) and high-field phases (phase III, IV) that appear in $H$~$\parallel$~$c$ configuration are similar to the $Y$- and $V$-phases, respectively, as expected for the classical 2$d$ TL Heisenberg AFM system.
The splitting of these phases occur due to additional interactions, such as NN interlayer interactions \cite{Yamamoto_RN2261}.
While the exact magnetic structure of these phases are not studied to date, we infer them by taking analogy from other 2$d$ TL Heisenberg AFM systems with similar features.
We note that Rb$_4$Mn(MoO$_4$)$_3$, a well-studied classical 2$d$ TL Heisenberg AFM system with easy-axis anisotropy, does not show any sign of phase transitions inside both the $Y$- and $V$-phases.
Rather, splitting of the phases are observed in easy-plane anisotropy systems RbFe(MoO$_4$)$_2$ and Ba$_3$CoSb$_2$O$_6$ in which additional interactions, such as NNN intralayer ($J'$) and/or NN interlayer ($J_{\perp}$) interactions are present.
In RbFe(MoO$_4$)$_2$, the low-field ground state is a 120~$\degree$ structure that modulates with an incommensurate (IC) wave vector along the $c$ axis.
It evolves to a $Y$-phase with magnetic field applied along the easy-plane direction through a first-order phase transition \cite{Svistov_RN2274,Sakhratov_RN2272}.
RbFe(MoO$_4$)$_2$ shows another first-order phase transition within the $V$-phase: a ``fan''-phase appears close to saturation magnetic field, which is a variation of the $V$-phase with IC modulation between the neighboring layers \cite{White_RN686,Sakhratov_RN2272,Zelenskiy_RN2282}.
The easy-plane anisotropic system Ba$_3$CoSb$_2$O$_6$ shows a first-order phase transition within the $V$-phase at $H$~$\sim$~0.7~$H_{sat.}$ that display a sharp peak in d$M$/d$H$ curve that signals the onset of another phase ($V'$) \cite{Susuki_RN1550}.

Thus, taking analogy with the RbFe(MoO$_4$)$_2$ case, we assume that phases I and II in NBMnPO both form 120~$\degree$ structures within the layers but differ in magnetic ordering pattern along the $c$ axis.
This idea is partially supported by the weak phase transitions between I$'$, II$'$, and II$''$ in the $H$~$\parallel$~$ab$ configuration (Fig.~\ref{Fig6}(b)).
If adjacent layers share the same 120~$\degree$ pattern at zero field, with in-plane magnetic fields, the system would form an inverted $Y$ structure having one of the spins pointing towards the magnetic field direction and evolve continuously to a fully polarized state by canting of spins without any phase transitions.
Phases III and IV can be considered as $V$- and its variant ($V'$), respectively.
Indeed further studies, such as neutron diffraction, are necessary to exactly understand the phases diagram of NBMnPO.

But the model including the NNN intralayer ($J'$) and/or NN interlayer ($J_{\perp}$) interactions fails to capture the emergence of a wide hysteresis region (colored region in Fig.~\ref{Fig6}(a)) in NBMnPO, which manifests as a hump in the up-sweep in $\chi'$($H$) curves, adjoining the $\uparrow$$\uparrow$$\downarrow$-phase on the high field side, in $H$~$\parallel$~$c$ configuration.

A plausible scenario for the hump region can be deduced from numerical simulation works by Seabra $et~al.$ \cite{Seabra_RN736,Seabra_RN2193}, where they studied the classical 2$d$ TL AFM system by including additional terms $J'$, $J_{\perp}$, and single-ion anisotropy to explain the complex low-field phase diagram, including a ``spin-supersolid'' phase \cite{Liu_RN2279}, observed in a metallic 2$d$ TL AFM system AgNiO$_2$ \cite{Coldea_RN2227}.
For a certain set of exchange interaction and single-ion anisotropy parameters, they obtained a phase diagram that consists of first-order phase transitions and phase coexistence regions adjacent to the $\uparrow$$\uparrow$$\downarrow$-phase (for example, see Fig.~24 in \cite{Seabra_RN2193}).
But this model relies on relatively strong next-nearest-neighbor and interlayer interactions ($J'$/$J$~=~0.15, $J_{\perp}$/$J$~=~-0.15) and single-ion anisotropy ($D$/$J$~=~0.5) that stabilizes a stripe phase at $H$~=~0 and a half-magnetization plateau under applied magnetic field which is absent in NBMnPO.

We note that the magnetic dipolar interaction in NBMnPO is non-negligible due to large magnetic moment of Mn$^{2+}$.
By using the magnetic moment of Mn$^{2+}$ obtained from CW analysis (5.60~$\mu_B$), the dipolar interaction energy in NBMnPO between NN, interlayer NNN, and intralayer NNN spins are calculated as 0.13~K, 0.024~K, and 0.055~K, respectively.
These values are comparable to $J$~$\sim$~0.43~K, obtained from mean-field theory, and may play an important role in understanding its unusual phase diagram.

Finally, we compare the phase diagrams of NB$M$PO series ($M$~=~Mn, Co, and Ni).
All three systems exhibit easy-axis magnetic anisotropy with 120~$\degree$ phase as the zero field state and, by application of $H$ along $c$ axis, evolve to $\uparrow$$\uparrow$$\downarrow$-, $V$-, and $V'$-phases.
The shapes of the phase diagrams are different in the classical (NBMnPO) and quantum spin (NBCoPO) limit.
In NBMnPO, phase boundaries between II--$\uparrow$$\uparrow$$\downarrow$, III--IV, and IV--PM display strong temperature dependences.
Especially, the $\uparrow$$\uparrow$$\downarrow$-phase region shrinks upon cooling, indicating that it is a thermally stabilized state.
On the other hand, phase boundaries of NBCoPO are nearly constant in $T$ \cite{Li_RN2099}.
This difference clearly manifests the role of thermal and quantum fluctuation in stabilizing the ordered phases in 2$d$ TL AFM systems.

\section{Conclusion}
Our experimental study on single crystals of NBMnPO shows that it display salient features of the classical 2$d$ TL Heisenberg AFM with easy-axis anisotropy, including the $\uparrow$$\uparrow$$\downarrow$-phase with 1/3 magnetization plateau when magnetic field is applied along the $c$ axis.
A detailed study of AC magnetic susceptibility reveals a unique hysteresis hump feature between the $\uparrow$$\uparrow$$\downarrow$ and $V$-phases as well as weak phase transitions.
While the weak phase transitions can be understood in terms of small additional NNN intralayer and/or NN interlayer interactions, the hysteresis hump require at least an order of magnitude large interaction scale and warrants further theoretical and experimental works in the future.
Our findings may also trigger to establish a model to understand the complicated QSL-like behavior and complex phase diagrams observed in sibling compounds NBCoPO and NBNiPO.

\ack
We thank T. H. Jang at MPK-POSTECH for carrying out the $C_p$ measurement in DR.
JK acknowledge Sungdae Ji, Bongjae Kim, Ara Go, Heung-Sik Kim, and Jiyeon Kim for helpful discussions and M. Kim for critical reading of the manuscript.
This work is supported by the KAERI Internal R\&D program (Grant nos. 524460-22, 79772-21) and National Research Foundation of Korea (No. 2021M2E3A3040092).
The National High Magnetic Field Laboratory is supported by the National Science Foundation through NSF/DMR-1644779 and the State of Florida.

\section*{References}
\bibliographystyle{iopart-num}

\begin{thebibliography}{100}

\bibitem{Collins_RN1353}
Collins M~F and Petrenko O~A 1997 {\em Can. J. of Phys.\/} {\bf 75} 605--655

\bibitem{Ramirez_RN1273}
Ramirez A~P 1994 {\em Ann. Rev. Mater. Sci.\/} {\bf 24} 453--480

\bibitem{Starykh_RN1367}
Starykh O~A 2015 {\em Rep. Prog. Phys.\/} {\bf 78} 052502

\bibitem{Wannier_RN1316}
Wannier G 1950 {\em Phys. Rev.\/} {\bf 79} 357--364

\bibitem{Husimi_RN1337}
Husimi K and Syozi I 1950 {\em Prog. Theor. Phys.\/} {\bf 5} 177--186

\bibitem{Anderson_RN1368}
Anderson P~W 1973 {\em Mater. Res. Bull.\/} {\bf 8} 153--160

\bibitem{Bernu_RN1421}
Bernu B, Lhuillier C and Pierre L 1992 {\em Phys. Rev. Lett.\/} {\bf 69}
  2590--2593
  
\bibitem{Capriotti_RN2135}
Capriotti L, Trumper A~E and Sorella S 1999 {\em Phys. Rev. Lett.\/} {\bf 82}
  3899--3902

\bibitem{White_RN1506}
White S~R and Chernyshev A~L 2007 {\em Phys. Rev. Lett.\/} {\bf 99} 127004

\bibitem{Balents_RN1917}
Balents L 2010 {\em Nature\/} {\bf 464} 199--208

\bibitem{Mila_RN1431}
Mila F 2000 {\em Eur. J. Phys.\/} {\bf 21} 499--510

\bibitem{Savary_RN1472}
Savary L and Balents L 2017 {\em Rep. Prog. Phys.\/} {\bf 80} 016502

\bibitem{Zhou_RN1537}
Zhou Y, Kanoda K and Ng T~K 2017 {\em Rev. Mod. Phys.\/} {\bf 89} 025003

\bibitem{Bramwell_RN1398}
Bramwell S~T and Gingras M~J 2001 {\em Science\/} {\bf 294} 1495--501

\bibitem{Gardner_RN1449}
Gardner J~S, Gingras M~J~P and Greedan J~E 2010 {\em Rev. Mod. Phys.\/} {\bf
  82} 53--107

\bibitem{Norman_RN1482}
Norman M~R 2016 {\em Rev. Mod. Phys.\/} {\bf 88} 041002

\bibitem{Chubukov_RN2252}
Chubukov A~V and Golosov D~I 1991 {\em J. Phys. Condens. Matter\/} {\bf 3}
  69--82

\bibitem{Kawamura_RN2239}
Kawamura H and Miyashita S 1985 {\em J. Phys. Soc. Jpn.\/} {\bf 54} 4530--4538

\bibitem{Seabra_RN1350}
Seabra L, Momoi T, Sindzingre P and Shannon N 2011 {\em Phys. Rev. B\/} {\bf
  84} 214418

\bibitem{Gvozdikova_RN2238}
Gvozdikova M~V, Melchy P~E and Zhitomirsky M~E 2011 {\em J. Phys. Condens.
  Matter\/} {\bf 23} 164209

\bibitem{Kitazawa_RN2215}
Kitazawa H, Suzuki H, Abe H, Tang J and Kido G 1999 {\em Phys. B\/} {\bf
  259-261} 890--891

\bibitem{Svistov_RN2201}
Svistov L~E, Smirnov A~I, Prozorova L~A, Petrenko O~A, Demianets L~N and
  Shapiro A~Y 2003 {\em Phys. Rev. B\/} {\bf 67} 094434

\bibitem{Ishii_RN1386}
Ishii R, Tanaka S, Onuma K, Nambu Y, Tokunaga M, Sakakibara T, Kawashima N,
  Maeno Y, Broholm C, Gautreaux D~P, Chan J~Y and Nakatsuji S 2011 {\em
  Europhys. Lett.\/} {\bf 94} 17001

\bibitem{Zhou_RN1657}
Zhou H~D, Xu C, Hallas A~M, Silverstein H~J, Wiebe C~R, Umegaki I, Yan J~Q,
  Murphy T~P, Park J~H, Qiu Y, Copley J~R~D, Gardner J~S and Takano Y 2012 {\em
  Phys. Rev. Lett.\/} {\bf 109} 267206

\bibitem{Quirion_RN2269}
Quirion G, Plumer M~L, Petrenko O~A, Balakrishnan G and Proust C 2009 {\em
  Phys. Rev. B\/} {\bf 80} 064420

\bibitem{Alicea_RN2256}
Alicea J, Chubukov A~V and Starykh O~A 2009 {\em Phys. Rev. Lett.\/} {\bf 102}
  137201

\bibitem{Takagi_RN2257}
Takagi T and Mekata M 1995 {\em J. Phys. Soc. Jpn.\/} {\bf 64} 4609--4627

\bibitem{Swanson_RN2258}
Swanson M, Haraldsen J~T and Fishman R~S 2009 {\em Phys. Rev. B\/} {\bf 79}
  184413

\bibitem{Yamamoto_RN2261}
Yamamoto D, Marmorini G and Danshita I 2015 {\em Phys. Rev. Lett.\/} {\bf 114}
  027201

\bibitem{Zhong_RN1618}
Zhong R, Guo S, Xu G, Xu Z and Cava R~J 2019 {\em Proc. Natl. Acad. Sci. U. S.
  A.\/} {\bf 116} 14505--14510

\bibitem{Li_RN2099}
Li N, Huang Q, Yue X~Y, Chu W~J, Chen Q, Choi E~S, Zhao X, Zhou H~D and Sun X~F
  2020 {\em Nat. Commun.\/} {\bf 11} 4216

\bibitem{Ding_RN2105}
Ding F, Ma Y, Gong X, Hu D, Zhao J, Li L, Zheng H, Zhang Y, Yu Y, Zhang L, Zhao
  F and Pan B 2021 {\em Chin. Phys. B\/} {\bf 30} 117505

\bibitem{Li_RN2117}
Li N, Huang Q, Brassington A, Yue X~Y, Chu W~J, Guang S~K, Zhou X~H, Gao P,
  Feng E~X, Cao H~B, Choi E~S, Sun Y, Li Q~J, Zhao X, Zhou H~D and Sun X~F 2021
  {\em Phys. Rev. B\/} {\bf 104} 104403

\bibitem{Sheldrick_RN2276}
Sheldrick G~M 2015 {\em Acta Cryst. C\/} {\bf 71} 3--8

\bibitem{Nishio-Hamane_RN2104}
Nishio-Hamane D, Minakawa T and Okada H 2014 {\em J. Mineral. Petrol. Sci.\/}
  {\bf 109} 34--37

\bibitem{Nenert_RN2103}
Nénert, Murshed, Hamed B, Gesing and Amara B 2020 {\em Z. Kristallogr. Cryst.
  Mater.\/} {\bf 235} 433--437

\bibitem{Supple}
{See Supplemental Material at XXX for the single crystal X-ray diffraction
  fitting parameters, refined crystal structure, and frequency-dependence of
  $\chi'$($H$).}

\bibitem{Yahne_RN2347}
Yahne D R, Pereira D, Jaubert L D C, Sanjeewa L D, Powell M, Kolis J W, Xu Guangyong, Enjalran M, Gingras M J P and Ross K A 2021 {\em Phys. Rev. Lett.\/} {\bf 127} 277206

\bibitem{Boukhris_RN2098}
Boukhris A, Hidouri M, Glorieux B and Amara M~B 2013 {\em J. Rare Earth.\/}
  {\bf 31} 849--856

\bibitem{Svistov_RN2274}
Svistov L~E, Prozorova L~A, Büttgen N, Shapiro A~Y and Demianets L~N 2005 {\em
  J. Exp. Theor. Phys.\/} {\bf 81} 102--107

\bibitem{Sakhratov_RN2272}
Sakhratov Y~A, Prokhnenko O, Shapiro A~Y, Zhou H~D, Svistov L~E, Reyes A~P and
  Petrenko O~A 2022 {\em Phys. Rev. B\/} {\bf 105} 014431

\bibitem{White_RN686}
White J~S, Niedermayer C, Gasparovic G, Broholm C, Park J~M~S, Shapiro A~Y,
  Demianets L~A and Kenzelmann M 2013 {\em Phys. Rev. B\/} {\bf 88} 060409(R)

\bibitem{Zelenskiy_RN2282}
Zelenskiy A, Quilliam J~A, Shapiro A~Y and Quirion G 2021 {\em Physical Review
  B\/} {\bf 103} 224422

\bibitem{Susuki_RN1550}
Susuki T, Kurita N, Tanaka T, Nojiri H, Matsuo A, Kindo K and Tanaka H 2013
  {\em Phys Rev Lett\/} {\bf 110} 267201

\bibitem{Seabra_RN736}
Seabra L and Shannon N 2010 {\em Phys. Rev. Lett.\/} {\bf 104} 237205

\bibitem{Seabra_RN2193}
Seabra L and Shannon N 2011 {\em Phys. Rev. B\/} {\bf 83} 134412

\bibitem{Liu_RN2279}
Liu K~S and Fisher M~E 1973 {\em J. Low Temp. Phys.\/} {\bf 10} 655--683

\bibitem{Coldea_RN2227}
Coldea A~I, Seabra L, McCollam A, Carrington A, Malone L, Bangura A~F,
  Vignolles D, van Rhee P~G, McDonald R~D, Sorgel T, Jansen M, Shannon N and
  Coldea R 2014 {\em Phys. Rev. B\/} {\bf 90} 020401(R)

\end{thebibliography}

\end{document}